%% file: main.tex
\documentclass{egpubl}
\usepackage{amsmath}
\usepackage{amssymb}
\usepackage{tabu}
\usepackage{booktabs}
\usepackage{lipsum}
\usepackage{mwe}
\usepackage{mathptmx} 
\usepackage{multirow}
\usepackage{multicol}
\usepackage{algorithm2e}
\usepackage{float}
\usepackage{xcolor}
\usepackage{enumitem}
\usepackage{thmtools,thm-restate}
\usepackage{tablefootnote}
\usepackage{makecell}
\usepackage{wrapfig}

\definecolor{mynavy}{rgb}{0.259, 0.490, 0.616}
\definecolor{mygray}{rgb}{0.439, 0.384, 0.2}

\SetKwComment{Comment}{/* }{ */}
\RestyleAlgo{ruled}

\BibtexOrBiblatex
\electronicVersion
\PrintedOrElectronic
\CGFccby
\usepackage[T1]{fontenc}
\usepackage{cite}
\usepackage{cleveref}
\title[A Prediction-Traversal Approach for Compressing Scientific Data on Unstructured Meshes with Bounded Error]
      {A Prediction-Traversal Approach for Compressing Scientific Data on Unstructured Meshes with Bounded Error}

\author[C. Ren, X. Liang, \& H. Guo]
{\parbox{\textwidth}{\centering Congrong Ren$^{1}$\orcid{0009-0006-6285-7271},
        Xin Liang$^{2}$\orcid{0000-0002-0630-1600},
        and Hanqi Guo$^{1}$\orcid{0000-0001-7776-1834} 
        }
        \\
{\parbox{\textwidth}{\centering $^1$Department of Computer Science and Engineering, The Ohio State University, USA\\
         $^2$Department of Computer Science, University of Kentucky, USA
       }
}}

\begin{document}
\maketitle
\begin{abstract}
We explore an error-bounded lossy compression approach for reducing scientific data associated with 2D/3D unstructured meshes. While existing lossy compressors offer a high compression ratio with bounded error for regular grid data, methodologies tailored for unstructured mesh data are lacking; for example, one can compress nodal data as 1D arrays, neglecting the spatial coherency of the mesh nodes. Inspired by the SZ compressor, which predicts and quantizes values in a multidimensional array, we dynamically reorganize nodal data into sequences.  Each sequence starts with a seed cell; based on a predefined traversal order, the next cell is added to the sequence if the current cell can predict and quantize the nodal data in the next cell with the given error bound.  As a result, one can efficiently compress the quantized nodal data in each sequence until all mesh nodes are traversed.
This paper also introduces a suite of novel error metrics, namely continuous mean squared error (CMSE) and continuous peak signal-to-noise ratio (CPSNR), to assess compression results for unstructured mesh data.  The continuous error metrics are defined by integrating the error function on all cells, providing objective statistics across nonuniformly distributed nodes/cells in the mesh.  We evaluate our methods with several scientific simulations ranging from ocean-climate models and computational fluid dynamics simulations with both traditional and continuous error metrics.  We demonstrated superior compression ratios and quality than existing lossy compressors.
\printccsdesc
\end{abstract}
\input{sec-intro}
\input{sec-related}
\input{sec-preliminaries}
\input{sec-method}
\input{sec-cmse}
\input{sec-evaluation}
\input{sec-discussion}
\input{sec-conclusions}

\section*{Acknowledgement}
This research is supported by the U.S. Department of Energy, Office of Advanced Scientific Computing Research (DE-SC0022753) and the National Science Foundation (OAC-2311878, OAC-2313123, IIS-1955764, OAC-2330367, and OAC-2313122). 

\bibliographystyle{eg-alpha-doi}
\bibliography{references}

\end{document}

%% file: sec-intro.tex
\section{Introduction}
\label{sec:intro}

Scientific data, usually characterized by their ever-increasing sizes with the growth of high-performance computing resources, cause grand challenges for scientists in storing, transferring, and understanding them. In recent years, \textit{error-bounded lossy compression}, or simply lossy compression, advanced significantly to reduce data size while maintaining an acceptable level of information fidelity. Lossy compression, thus, has been treated as a methodology to alleviate the pressure caused by large-scale scientific data. Visualization tasks such as volume rendering~\cite{chow1997optimized,schneider2003compression}, isosurfacing~\cite{li2017spatiotemporal}, and topological data analysis~\cite{soler2018topologically,yan2023toposz} have greatly benefited from lossy compression.

To date, the community's primary focus for lossy compression is regular grid data, where values are stored in an implicit order representing 2D and 3D domains.
In our observation, most existing algorithms rely on a block/stencil (e.g., $4^3$ block in the orthogonal transformation in ZFP~\mbox{\cite{lindstrom2014fixed}} and $2^3$ stencil in the Lorenzo predictor in SZ~\mbox{\cite{tao2017significantly}}) to compress regular grid data. However, efficient compression of \emph{unstructured mesh data} is lacking, which is more challenging due to the lack of stencil structure for unstructured mesh data and the requirement on storing both nodal data and mesh information. Vertices in an unstructured mesh are oftentimes arbitrarily indexed in memory, oblivious to spatial coherency among vertices, leading to a random data layout that challenges compressors. For example, a previous study~\mbox{\cite{LiangDCRLOCPG23}} compressed nodal data as a 1D series with a limited compression ratio in a CFD simulation. Furthermore, unlike regular grids, unstructured mesh data are composed of not only the values on vertices but also the mesh's topology. This necessitates considering two distinct aspects (mesh and nodal data) during compression and decompression.

We motivate this work by the amount of time-varying and multivariate data associated with a static mesh. For example, in a Model for Prediction Across Scales-Ocean (MPAS-Ocean)~\mbox{\cite{e3sm}} simulation, a 60-year of monthly dataset across 80 depth layers with 115 variables, including temperature, salinity, velocity, and pressure, share the same underlying unstructured mesh.
In this case, the aggregated size of the nodal data is $O(10^6)$ larger than the mesh itself and necessitates the reduction of the nodal data.

To these ends, this paper introduces a novel error-bounded lossy compression method for 2D/3D unstructured mesh data.
Inspired by the SZ~\cite{tao2017significantly} compressor, which predicts and quantizes values in a multidimensional array, we dynamically reorganize nodal data into sequences. The key innovation compared with SZ is our tailored predictor for unstructured meshes: we sequentially visit, predict by extrapolation, and quantize the predicted value of every mesh node, where the order of visiting mesh nodes is determined by traversing on the dual graph of the unstructured mesh. Each sequence starts with a seed cell; based on a predefined traversal order, the next cell is added to the sequence if the current cell can predict and quantize the nodal data in the next cell within the given error bound. The sequence grows until an unpredictable node is met, and a new sequence is initiated until all nodes are traversed. As a result, one can efficiently compress the quantized nodal data in each sequence until all mesh nodes are traversed.

To measure the quality of interpolated value preservation on unstructured meshes of continuous fields, we introduce a new metric, \textit{continuous mean squared error} (CMSE), which is defined by generalizing a traditional metric, MSE, that accumulates pointwise errors to a metric that aggregates cellwise errors. CMSE measures the overall statistical distortion of decompressed data on the continuous domain instead of on mesh nodes only and is straightforward to generalize to other metrics, including root MSE (RMSE), normalized RMSE (NRMSE), and PSNR for continuous domains. We also demonstrate why CMSE is more meaningful than traditional pointwise difference metrics when measuring data distortion in a continuous domain by several scenarios and evaluate our algorithm on both CMSE and pointwise error metrics.  In summary, this paper makes the following contributions:
\begin{itemize}
    \item A novel algorithm for data compression on 2D and 3D unstructured meshes;
    \item A comprehensive evaluation of our compression method with our proposed continuous error metrics.
\end{itemize}

%% file: sec-related.tex
\section{Related Work}
\label{sec:related}

We summarize related work in error-bounded lossy compression and mesh compression. 

\subsection{Error-bounded Lossy Compression for Scientific Data}

One can categorize compression algorithms into lossless and lossy compression. Lossless compression achieves compression by eliminating redundancy within the data while allowing complete reconstruction of the original data upon decompression, while lossy compression keeps an acceptable amount of information and largely reduces data size.  However, lossless compressors are usually unsuitable for scientific data due to their relatively low compression ratios (generally not much better than 2:1) for floating-point numbers~\cite{zhao2021optimizing}.  For scientific data, lossy compression usually delivers much higher compression ratios with quality guaranteed.  One can further categorize lossy compression into error-bounded and non-error-bounded lossy compression based on whether the pointwise error is restricted in user-specified error bounds.

Almost all existing lossy compressors are specific to regular grid data.  For example, one can compress regular-grid data with prediction-based, transform-based, dimension-reduction-based, and neural-based methods. 
Prediction-based compressors, such as ISABELA~\mbox{\cite{lakshminarasimhan2011compressing}}, FPZIP~\mbox{\cite{lindstrom2006fast}}, MGARD~\mbox{\cite{ainsworth2020multilevel}}, and SZ family~\mbox{\cite{liang2018error,liang2022sz3,tao2017significantly,zhao2021optimizing,zhao2020significantly}}, apply predictors (e.g., Lorenzo predictor~\cite{ibarria2003out}) first to estimate the values of unknown data points according to known information such as predicted data points, and then to use a quantization scheme, e.g., relative coordinate quantization~\mbox{\cite{rusinkiewicz2000qsplat}}, to limit pointwise error into a user-specified bound. Transform-based compressors (e.g., ZFP~\cite{lindstrom2014fixed} and wavelet~\cite{LiSOMCC17}) transform original data into sparsely-distributed coefficients that are easier to compress.  Dimension-reduction-based compressors (e.g., TTHRESH~\cite{ballester2019tthresh}) reduce data dimensions by techniques such as higher-order singular vector decomposition (HOSVD).  Recently, neural networks have been widely used to reconstruct scientific data, such as autoencoders~\cite{liu2021exploring, ZhangGSWP22}, superresolution networks~\cite{WursterGSPX23, Han-VIS21}, and implicit neural representations~\cite{INR22, Lu21, WeissHW22, Martel21, Sitzmann-NIPS20}. Yet, most neural compressors do not offer explicit pointwise error control for scientific applications.

Few error-bounded lossy compressors consider unstructured mesh data. Iverson et al.~\cite{Iverson12} proposed a domain-decomposition approach that divides the mesh into regions first, followed by representing each region using the mean value of the nodal data.  The main drawback is the high storage overhead of the regions represented as integer arrays, making it impractical for large-scale data. Researchers also explored non-error-bounded compression of 2D unstructured grid data.  For example, Kamath~\mbox{\cite{kamath2020compressing}} proposed a training-based approach that targets a fusion plasma simulation code with 2D triangular grids. Yet, the method requires extra handling of high-error points with a regression algorithm. Salloum et al.~\mbox{\cite{salloum2018optimal}} used a compressed sensing approach to compress 2D unstructured grid data, which requires an iterative and optimization process to decompress the data.  Salloum et al.~\mbox{\cite{salloum2020comparing}} further explored using Alpert multi-wavelets for compressing unstructured grid data. Another related research is the compression of adaptive mesh refinement (AMR) data~\cite{WangPGTJTSD0FLC23}, which relies on AMR's semi-structured blocks.

\subsection{Unstructured Mesh Compression}

Although related, the objective of our research is fundamentally different from compressing the mesh itself see Peng et al.~\mbox{\cite{Peng05}} for a comprehensive review). Instead, we assume that one can afford to store an accurate mesh, and the data reduction challenge stems from the accumulation of time-varying and multivariate data.

Mesh simplification offers a lossy manner to reduce a mesh with edge-collapsing~\cite{uesu2005simplification}, vertex clustering~\cite{rossignac1993multi}, and wavelet-based methods~\cite{gross1996efficient}.  For a review of surface mesh simplification methods, see Cignoni et al.~\cite{cignoni1998comparison}.  With edge-collapsing, one can define different edge-cost functions serving different feature-preserving goals~\cite{uesu2005simplification}. For example, Natarajan and Edelsbrunner~\cite{natarajan2004simplification} preserved the density map of mesh by a metric of angles; Tram et al.~\cite{cignoni2000simplification} preserved mesh boundary by controlling boundary constraints; Chiang and Lu~\cite{chiang2003progressive} preserved topology of isosurfaces by introducing multiple levels of details of a tetrahedral mesh. Vertex clustering methods group nearby vertices into clusters based on various criteria, such as geometric proximity or attribute similarity. Each cluster is then replaced by a new representative vertex in a simplified mesh. Vertex clustering is fast but makes the geometry or topology of the original mesh hard to preserve. Wavelet-based methods apply wavelet transform to analyze the level of details needed by a local region. However, a regular, hierarchical decomposition of the surface is required to enable wavelet decomposition. 

Besides, our method relates to but fundamentally differs from triangle strips algorithms~\mbox{\cite{estkowski2002optimal}}, which compute an optimal decomposition of triangle strips from a polygonal surface model with a given cost function.  While one could potentially formulate our compression outputs as triangle strips, our method does not need to adhere to clockwise or counter-clockwise alternation.
Also, triangle strips algorithms generally have high complexity and are difficult to generalize to tetrahedral grids for compression.

%% file: sec-preliminaries.tex
\section{Preliminaries}
\label{sec:preliminaries}

This section formalizes the inputs of our algorithm and reviews three preliminaries: SZ compressors~\mbox{\cite{tao2017significantly}}, barycentric interpolation, and evaluation metrics for lossy compression.

\subsection{Unstructured meshes and unstructured mesh data}
\label{sec:problem_formulation}

Without loss of generality, we consider 2D/3D simplicial meshes, which contain only triangular/tetrahedral cells with a piecewise linear interpolation basis.  If the input mesh is nonsimplicial and contains higher-order elements, one can still use our method with a triangulated mesh~\cite{de2010triangulations, si2019simple} without adding new vertices.  In this case, vertices in the original mesh remain in the exact ordering, and our algorithm can still guarantee bounded error for the variable values on each vertex, as described later.

We assume that the input mesh is constant for all variables across all timesteps of the dataset.  Considering a scalar variable $f: \mathbb{X}\rightarrow\mathbb{R}$, where $\mathbb{X}\subset\mathbb{R}^2$ or $\mathbb{R}^3$ is the domain defined by the mesh complex $M=\langle V, C \rangle$, $f$ is represented by the \emph{nodal values} $\{f_i\}, i\in V$, where $V$ and $C$ are the sets of vertices and cells, respectively.  The lossily compressed $f$ is defined by a user-specified \emph{(absolute) error bound} $\xi$ such that $|f_i-\hat{f}_i|\leq\xi$ for all $i\in V$, where $\hat{f}_i$ denotes the \emph{decompressed value} of $f_i$. 

\subsection{Prediction-based lossy compression}
\label{sec:sz1.4}

For a comprehensive picture of our method, which essentially redesigns the prediction stage of SZ~\cite{tao2017significantly}, we review the four major stages of prediction-based lossy compressors.  Our algorithm leverages existing modules provided by the SZ3 package~\cite{liang2022sz3}, which implements all necessary building blocks for prediction-based lossy compressors. In general, there are four major phases in prediction-based lossy compressors: prediction, quantization, encoding, and lossless compression, as reviewed below.

\textbf{Prediction}.  The prediction modules use interpolation or transformation to predict neighboring data values.  For example, SZ 1.4 uses the \textit{Lorenzo predictor}~\cite{ibarria2003out} to predict the value of an unvisited data point based on decompressed values of its neighbors.  If the prediction is within the error bound, the value is \emph{predictable} and further represented as a fixed-precision quantization code in the quantization module described below; otherwise, the value is \emph{unpredictable} and must be stored in a lossless manner.

\begin{figure}[!htb]
    \centering
    \includegraphics[width=\linewidth]{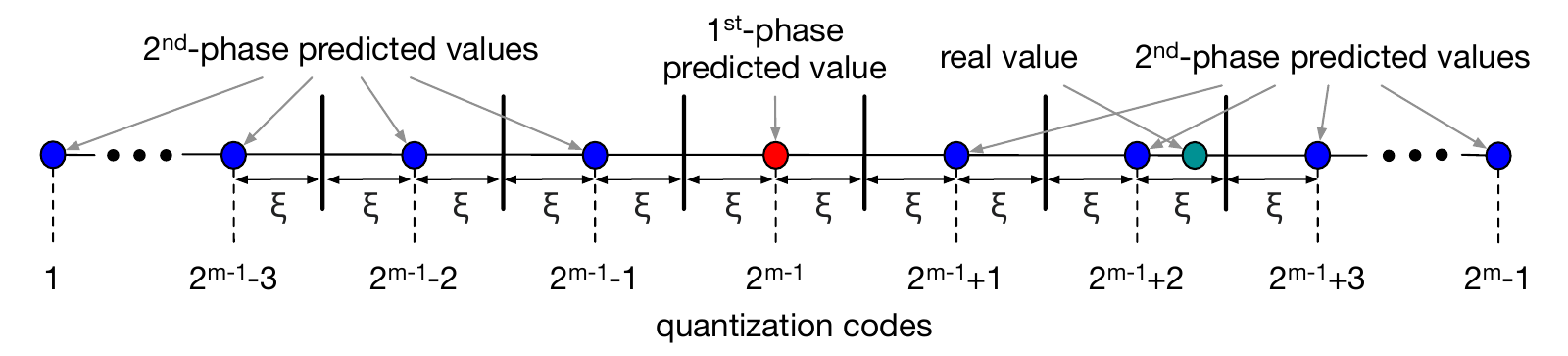}
    \caption{Linear-scaling quantization.  For the value in this case, the quantization code is $2^{m-1}+2$. Image reproduced from Figure 2 in Tao et al.~\cite{tao2017significantly}.}
    \label{fig:quantization}
\end{figure}

\textbf{Quantization}.  The quantization step uses an error-controlled quantization encoder (\Cref{fig:quantization}) to convert a predictable value to a quantization code. We use the linear-scale quantization~\cite{tao2017significantly} in this paper.  Let $m$ denote the number of bits (we set to be 16 to cover reasonable range of error bounds~\mbox{\cite{tao2017significantly}}) used to encode a data point; for the data value $x$, one can first obtain its predicted value (shown as the red dot); imagine a series of \textit{$2^{nd}$-phase predicted values} (shown as blue dots in~\Cref{fig:quantization}) that separate the 1D axis linearly together with $1^{st}$-phase predicted value into segments with length of $2\xi$, as well as a partition of the axis in which each interval takes two successive $1^{st}$- or $2^{nd}$-phase predicted values as midpoint.  We represent every interval by an integer, so-called \textit{quantization code}. The interval with $1^{st}$-phase predicted value as midpoint is represented by $2^{m-1}$, and other intervals are represented by consecutive integers by the order of their midpoints. There are $2^m-1$ intervals with quantization codes between $1$ and $2^m-1$; a data point lying within these intervals is predictable. 

\textbf{Encoding and lossless compression}.  With prediction and quantization, data are transformed into two sets as bitstreams: quantization codes (for predictable values) and floating point numbers (for unpredictable values).  For example, SZ 1.4 uses a customized Huffman encoder~\cite{huffman1952method} to transform the bitstreams into a compact representation. After encoding, one can further reduce the encoded data with off-the-shelf lossless compressors such as ZSTD~\cite{zstd}.

\begin{figure*}[!th]
    \centering
    \includegraphics[width=0.95\textwidth]{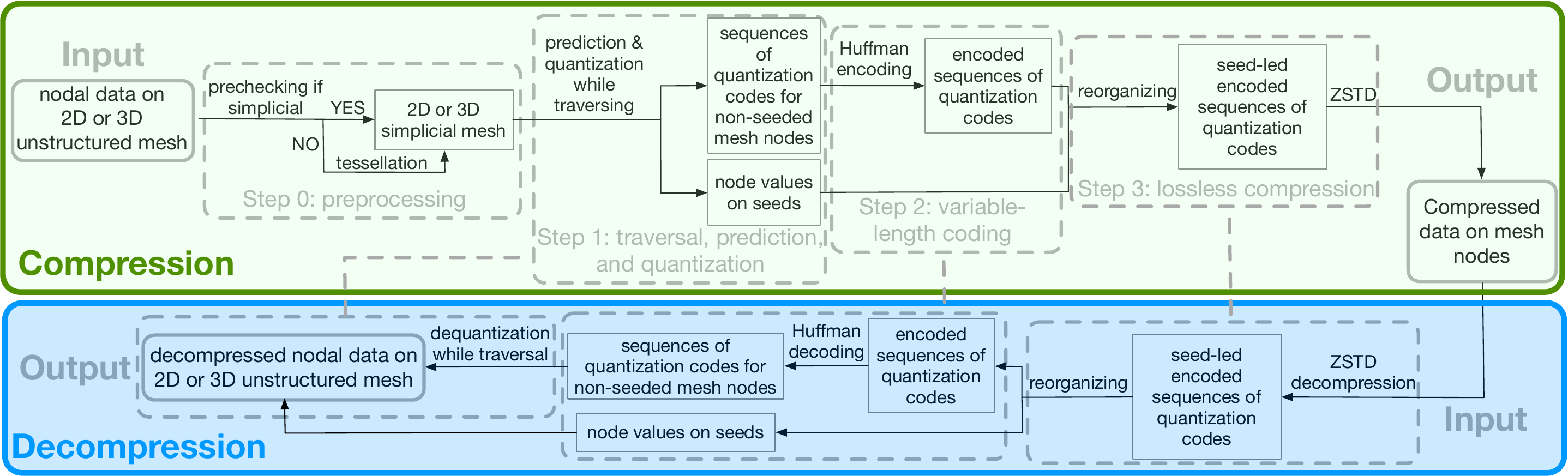}
    \caption{Workflow of our compression/decompression algorithm for unstructured mesh data.}
    \label{fig:workflow}
\end{figure*}

\subsection{Barycentric Interpolation and Extrapolation}
\label{sec:interp&extrap}

This research uses barycentric extrapolation as the basis of the prediction for unstructured mesh data. Still, the prediction requires a deterministic (and preferably implicit) vertex ordering, which we will detail in the next section. The barycentric interpolation scheme establishes the piecewise linear representation of a simplicial mesh.  We explain the interpolation scheme with 3D tetrahedral meshes, and the same scheme can be easily applied to 2D triangular meshes. For an arbitrary location $\mathbf{p}\in\mathbb{X}$, one can estimate $f(\mathbf{p})$ by first locating the cell $c\in C$ that contains $\mathbf{p}$, and then we have
\begin{equation}
    f(\mathbf{p})=\lambda_0f_0+\lambda_1f_1+\lambda_2f_2+\lambda_3f_3,
\end{equation}
where $\{\lambda_i\}$ are the \emph{barycentric coordinates} of $\mathbf{p}$ with respect to $c$, and $\{f_i\}$ are the scalar values on the four vertices of $c$, $i=0,1,2,3$.  With the barycentric coordinates, we also have $\mathbf{p}=\sum_{i=0}^3\lambda_i \mathbf{p}_i$ and $\sum_{i=0}^3\lambda_i=1$, where $\mathbf{p}_i=(x_i, y_i, z_i)^\intercal$ are the coordinates of vertex $i$.  Assuming the cell is nondegenerate, one can transform between Cartesian and barycentric coordinates with
\begin{equation}
    \left(x, y, z, 1\right)^\intercal =
    \left(\begin{smallmatrix}
        x_0 & x_1 & x_2 & x_3 \\
        y_0 & y_1 & y_2 & y_3 \\
        z_0 & z_1 & z_2 & z_3 \\
        1   & 1   & 1   & 1
    \end{smallmatrix}\right)
    \left(\lambda_0, \lambda_1, \lambda_2, \lambda_3\right)^\intercal.
    \label{eq:barycentric_coors}
\end{equation}
Barycentric extrapolation uses the same equations as above, except that the point to be predicted is outside of the cell. Our prediction scheme uses barycentric extrapolation for a point $\mathbf{p}^\prime$ outside a cell with the same formulation as the barycentric interpolation; more specifically, $\mathbf{p}^\prime$ is one of the neighboring cells' vertices, as further described in the next section. 

\subsection{Metrics for evaluating lossy compression results}
\label{traditional_matrics}

One can quantify how well the decompressed data retains essential information while adhering to the defined error thresholds.

\textbf{Size metrics} quantify how much a compressor can reduce the data size. Typical size metrics include \textit{compression ratio} (CR), the ratio of original data size to compressed data size, and \textit{bitrate} (BR), the average number of bits used to represent each value. 

\textbf{Error metrics} measure how the decompressed data deviates from the original data after compression, but all existing error metrics measure pointwise (in the context of unstructured meshes, vertexwise) error. Because an unstructured mesh represents the continuous domain $\mathbb{X}$, new metrics are needed to measure the compression quality of unstructured mesh data. Examples of vertexwise error metrics include MSE, which measures the average of squared differences between nodal data:
\begin{equation}
    \mathtt{MSE}(f,\hat{f})=1/|V|\sum\nolimits_{i\in V}(\hat{f}_i-f_i)^2.\label{eq:mse}
\end{equation}
The root mean squared error $\mathtt{RMSE}(f,\hat{f})=\sqrt{\mathtt{MSE}(f,\hat{f})}$ scales MSE to the original units of the data.  The normalized RMSE (NRMSE) further scales RMSE: $\mathtt{NRMSE}(f,\hat{f})=\mathtt{RMSE}(f,\hat{f})/ (f_{\max}-f_{\min})$, where $f_{\min}$ and $f_{\max}$ denote the value range of $f$.  One can further measure PSNR, the ratio between the maximum possible value of the data and the value of distorting noise that affects the quality of the data in a logarithmic decibel scale: $\mathtt{PSNR}=-20\log_{10}\frac{f_{\max}-f_{\min}}{\mathtt{RMSE}(f,\hat{f})}$.  We will further propose continuous versions of MSE-based metrics later in this paper.

%% file: sec-method.tex
\section{Our method}
\label{sec:method}

\Cref{fig:workflow} illustrates the workflow of our data compression and decompression methods for a simplicial mesh. We apply a graph traversal-based method to traverse the mesh and determine the sequence order of mesh nodes in both compression and decompression processes. Both prediction and quantization are involved in traversal, and we predict the values of newly visited mesh nodes by barycentric extrapolation.

\begin{figure*}[!th]
    \centering
    \includegraphics[width=0.98\textwidth]{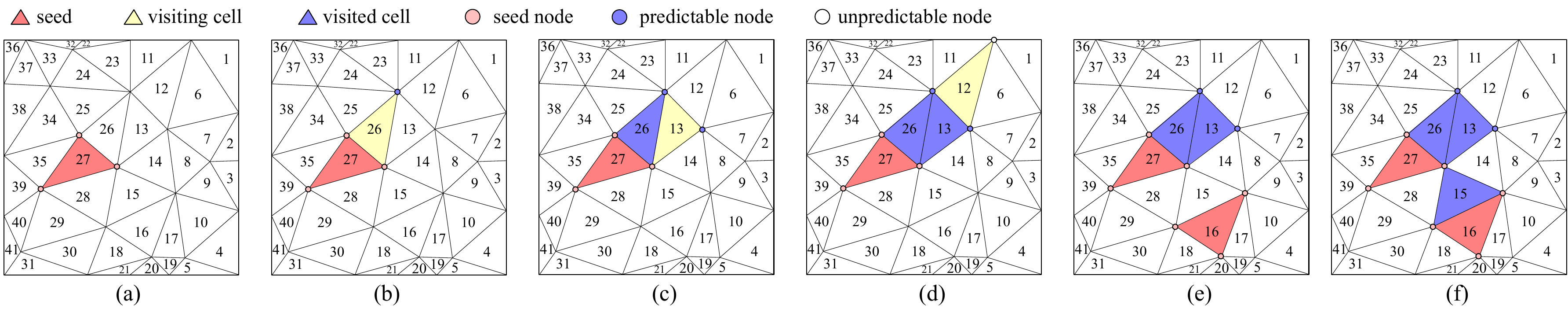}
    \caption{The process of traversal on the mesh. We assume that all the cell indices are in-memory ordered. (a) A cell is randomly selected to be ``seed'' (marked in red with index 27). The seed is directly marked as visited, and the values of all the nodes on the seed are losslessly stored. (b) The cell with the smallest index among the neighbors of the last visited cell (the seed in red at this step) is selected as the cell to visit next, marked in yellow and with index 26. The node newly introduced by the yellow cell is predicted by barycentric extrapolation w.r.t. the last visited cell and encoded by quantizer shown in~\Cref{fig:quantization}. We then determine whether this node is predictable. (c) If the newly visited node is predictable, we mark current cell \#26 as visited and repeat all the operations in (b). Then we visit cell \#13, predict and quantize its newly introduced node, and determine whether it is predictable. (d) We mark cell \#13 as visited and continue with its neighbor with the smallest index, cell \#12. After prediction and quantization, we determine whether it is unpredictable. (e) If the newly introduced node is unpredictable, we do not mark the current cell (i.e., cell \#12) as visited and instead terminate traversal starting with current seed, cell \# 34. We randomly select a new seed (cell \#16 marked in red) from the set of unvisited cells and repeat traversal. (f) The cell (\#15) with all visited nodes is marked as visited. The whole algorithm terminates when all nodes are visited.}
    \label{fig:traversal}
\end{figure*}

\subsection{Traversal, prediction, and quantization}

We use a greedy strategy to reorder the nodes into sequences that can be predicted and quantized.  The basic idea is to establish each sequence with a seed cell, and an adjacent cell is added if the value on the nonshared vertex can be predicted by the current cell.  As such, our method can leverage the spatial coherency of neighboring cells and vertices to predict values by extrapolation.  

Another key to our method is to use a pre-determined and implicit traversal order.  Specifically, we always visit the neighboring cell with the minimum cell index.  Such an implicit traversal order avoids the storage of cell indices in compressed data, significantly reducing the possibility of storage overhead because explicit traversal orders could occupy a large amount of storage space. The choice of traversal orders is further discussed and evaluated in~\mbox{\Cref{sec:ablation}}.

\begin{figure}
    \centering
    \includegraphics[width=\linewidth]{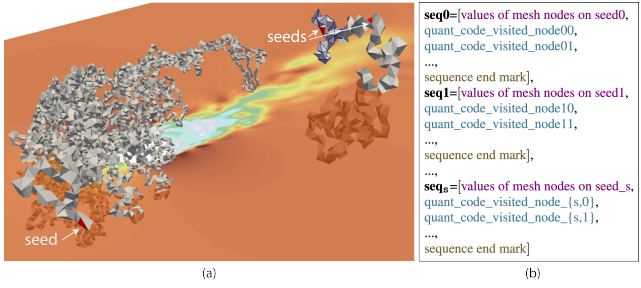}
    \caption{(a) Visualization of three traversal sequences in a 3D unstructured grid data (LES-s); the semi-transparent surface is a sliced plane of velocity magnitude visualized as context. (b) Data layout of traversal sequences.}
    \label{fig:traversal_fig}
    \label{fig:sequences}
\end{figure}

The pseudocode of our prediction-traversal-quantization algorithm is shown in Algorithms~\ref{alg:compress} and~\ref{alg:traverse}.  
As illustrated in \Cref{fig:traversal}, we work on the mesh where vertex indices determine the original memory layout.
We start with an arbitrary cell as the seed and losslessly store all nodal values of the seed.
Then, we iteratively increase the sequence length by traversing the mesh's dual graph; the dual graph's nodes are $C$, and two cells are connected if they have a shared face.
Every time we visit a new cell, we also get a newly visited node to be predicted in the cell. Then, we predict the value of the newly visited node. Multiple predictors can be applied in this stage with proper assumption of the data; for simplicity, we apply barycentric extrapolation in~\Cref{sec:interp&extrap} w.r.t. the last visited cell to predict the newly visited node. Comparison over different predictors is not covered in this work. The predicted value is transformed with the error-controlled quantization (\Cref{sec:sz1.4}).  The iteration terminates if (1) the newly visited node is unpredictable or (2) all the neighbors of the current visiting cell have been visited.
After the termination, if an unvisited node exists, we start a new sequence with an unvisited seed cell. Several sequences on a real-world 3D dataset in shown in~\mbox{\Cref{fig:sequences}} (a).

Our (de)compression algorithm takes $O(|V|)$ to terminate because all mesh nodes are visited once. Around 2\% of mesh nodes have been predicted before by other cells but are then included in later-chosen seed(s). Note that some cells will never be visited by cell traversal until the algorithm terminates if all their mesh nodes are visited via other cells.

\subsection{Quantized data layout, encoding, and compression}

With the prediction-traversal process, we organize quantization codes into sequences, as illustrated in~\Cref{fig:sequences} (b).  
Each sequence is led by the seed and followed by all visited cells and thus consists of three components: \textcolor{violet}{values of mesh nodes on a seed} are stored as a double precision floating point, \textcolor{mynavy}{quant\_code of visited\_node} and \textcolor{mygray}{sequence end mark} are stored as integers.
Because our traversal order is implicit, there is no need to store it.  
After all sequences are complete, we use SZ3's encoding and lossless compression routines to further reduce the sequences' size, as reviewed in~\Cref{sec:sz1.4}.

We analyze the storage cost of our data layout.  
Let $n_{seq}$ denote the number of sequences shown in the box of~\Cref{fig:traversal_fig}, equivalent to the number of seeds one chooses to visit all nodes. Let $d$ denote the dimension of the mesh. For double-precision data points, the numbers of nodes in sequences
in~\Cref{fig:traversal_fig} are within $[2|V|+(6d+4)n_{seq},2|V|+(8d+2)n_{seq}]$ bytes, which consists of three parts: the size of (1) values on all seeds is $8d\times n_{seq}$ because we losslessly store the values of seeds by 8 bytes; (2) all non-seeded values is within $[2(|V|-dn_{seq}),2(|V|-n_{seq})]$, because we have ($|V|-dn_{seq}$) non-seeded nodes if all the nodes of each seed are not visited by other sequences and ($|V|-n_{seq}$) non-seeded nodes if only one node of each seed are not visited by other sequences. 2 bytes are used for the quantization code of every node; (3) all end marks (all are single-precision infinity in our method) is $4 n_{seq}$.

\begin{algorithm}[!ht]
\caption{Compress}\label{alg:compress}
{\small
\KwData{mesh, nodal\_values}
seeds\_values $\gets$ []\;
quant\_codes $\gets$ []\;
decomp\_values $\gets$ zero vector with the same length as nodal\_values\;
\While{not all nodes are visited}{
seed $\gets$ a randomly chosen cell from unvisited cells\;
seed.\textit{visited} $\gets$ \texttt{True}\;
\For{node \texttt{in} seed.nodes}{
node.\textit{visited} $\gets$ \texttt{True}\;
decomp\_values[node] $\gets$ nodal\_values[node]\;
seeds\_values.\textit{append(}nodal\_values[node]\textit{)}\;
check if any unvisited cells incident to node have all visited vertices after node is visited. If yes, set the cell(s) to be visited\;
}
quant\_codes, decomp\_values $\gets$ TraverselFromSeed(seed, mesh, quant\_codes, decomp\_values)\;
}
quant\_codes $\gets$ Huffman encoding quant\_codes\;
sequences $\gets$ organize seeds\_values and quant\_codes\;
sequences $\gets$ lossless compress sequences by ZSTD\;
\textbf{return} sequences
}
\end{algorithm}

\begin{algorithm}[!ht]
\caption{TraversalFromSeed}\label{alg:traverse}
{\small
\KwData{seed, mesh, quant\_codes, decomp\_values}
stack $\gets$ [(seed, \texttt{None})]\;
\While{stack$!=\emptyset$}{
current\_cell, previous\_cell$ \gets$ stack.\textit{pop()}\;
\uIf{current\_cell.visited $==$ \texttt{False}}{
newly\_visited\_node $\gets$ set(current\_cell) - set(previous\_cell)\;
predict value of newly\_visited\_node by barycentric extrapolation w.r.t. the decomp\_values on previous\_cell\;
decompressed\_value, quant\_code $\gets$ quantize the predicted value\;
\eIf{newly\_visited\_node is predictable}{
quant\_codes.\textit{append(}quant\_code\textit{)}\;
decomp\_values[newly\_visited\_node] $\gets$ decompressed\_value\;
}{
quant\_codes.\textit{append(}end\_mark\textit{)}\;
\texttt{break}\;
}
newly\_visited\_node.\textit{visited} $\gets$ \texttt{True}\;
check if any unvisited cells incident to newly\_visited\_node have all visited nodes after newly\_visited\_node is visited. If yes, set the cell(s) to be visited\;
}
\textit{push(}(neighbor\_with\_min\_idx, current\_cell)\textit{)}
}
\textbf{return} quant\_codes, decomp\_values\;
}
\end{algorithm}

%% file: sec-cmse.tex
\begin{figure}[!th]
    \centering
    \includegraphics[width=\linewidth]{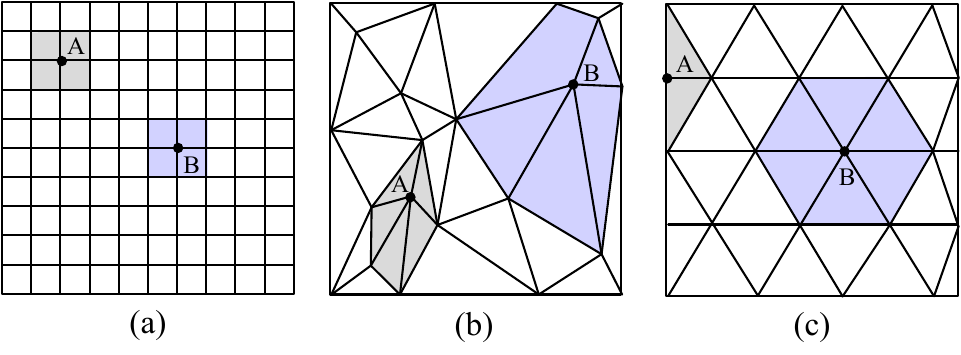}
    \caption{Illustration of regions (filled by gray or blue) affected by different mesh nodes in three meshes during interpolation. (a) In a regular grid, any two interior mesh nodes (e.g., \textit{A} and \textit{B}) affect the values of points in regions with the same size. (b) In an unstructured mesh, some points (e.g., \textit{A}) are incident to cells with smaller areas while some points (e.g., \textit{B}) affect larger areas. (c) Even if mesh nodes are uniformly distributed in the field, boundary nodes (e.g., \textit{A}) affect smaller regions than interior nodes (e.g., \textit{B}) do.}
    \label{fig:cmse_illustration}
\end{figure}

\section{Continuous Mean Squared Error (CMSE)}
\label{sec:cmse}

We propose a suite of novel continuous error metrics to evaluate the quality of lossy compressed data on unstructured meshes.
Metrics commonly used to measure compression quality, including MSE and its variants such as RMSE, NRMSE, and PSNR, are pointwise errors on only mesh nodes when applied on unstructured meshes and weigh all mesh nodes equally.  However, in a continuous domain, every position is associated with a function value, which shows the continuity of the underlying physical principles and thus needs taking into account~\mbox{\cite{bachthaler2008continuous, duffy2012integrating}}. Every mesh node affects interpolated values of all positions in the cells that are incident to the mesh node. Different than Salloum et al.~\mbox{\cite{salloum2020comparing}}, which compares the spectral amplitudes of two continuous fields after wavelet transformation, we generalize MSE to fields represented by unstructured meshes.

Traditional pointwise metrics are suitable for uniformly structured meshes, where the value on a mesh node contributes to a fixed area of the region (\Cref{fig:cmse_illustration} (a)). However, in a non-uniform grid or unstructured mesh, the total area of cells incident to a mesh node varies node by node. Some regions on the field are sampled sparsely, and thus mesh nodes in those regions are incident with larger cells (\Cref{fig:cmse_illustration} (b)); some nodes are on the boundary of the field and incident to fewer cells (\Cref{fig:cmse_illustration} (c)). To consider the error over all positions in a continuous field, we weigh the influence of mesh nodes according to how large the region in which they affect values on points by continuous mean squared error (CMSE).

Recall the definitions in~\Cref{sec:interp&extrap}, the pointwise squared error at a point $\mathbf{p}\in\mathbb{X}$ is defined by
$e(\mathbf{p})=(\hat{f}(\mathbf{p})-f(\mathbf{p}))^2$. We define \textit{cellwise squared error} of a tetrahedral cell $c$ to be the integral of pointwise squared error in the cell:
\begin{equation*}
    E(c)=\iiint_c e(p)dV,
\end{equation*}
where $dV$ is the infinitesimal volume of the cell. While MSE takes an average of pointwise squared error over all mesh nodes (\mbox{\Cref{eq:mse}}), CMSE takes an average of pointwise squared error of all points in a continuous field:
\begin{equation*}
    \texttt{CMSE}(f,\hat{f})=1/|V_C|\sum\nolimits_{i\in C}E(c_i),
\end{equation*}
where $V_C$ is the total volume of cells in the mesh and $c_i$ is the $i$th cell. With CMSE defined, we can define continuous RMSE, NRMSE, and PSNR based accordingly, as shown in~\mbox{\Cref{tab:continuous_metrics}}. The closed form of cellwise squared error varies by interpolation method. We take barycentric interpolation as an example. The pointwise squared error is
\begin{equation}
    e(\mathbf{p})=\left(\sum\nolimits_{i=0}^3\lambda_i(\hat{f}_i-f_i)\right)^2
    \mbox{,}
    \label{eq:ptws_se}
\end{equation}
where $\lambda_0$ to $\lambda_3$ are barycentric coordinates of $p$ w.r.t. the cell $c$ containing $\mathbf{p}$.
Cellwise squared error of cell $c$ is then given by:
\begin{align}
    E(c)=&\iiint_c e(p(x,y,z))dV\nonumber\\
    =&\int_0^1\int_0^{1-\lambda_0}\int_0^{1-\lambda_0-\lambda_1} e(p(\lambda_0,\lambda_1,\lambda_2))\cdot|J_c| d\lambda_2d\lambda_1d\lambda_0\mbox{,}\nonumber
\end{align}
where $J_c$ is the Jacobian of the mapping from $(x,y,z)$ to $(\lambda_0,\lambda_1,\lambda_2)$:
\begin{equation}
    J_c=\frac{\partial(x,y,z)}{\partial(\lambda_0,\lambda_1,\lambda_2)}=\left[
    \begin{smallmatrix}
         \frac{\partial x}{\partial\lambda_0} & \frac{\partial x}{\partial\lambda_1} & \frac{\partial x}{\partial\lambda_2} \\
         \frac{\partial y}{\partial\lambda_0} & \frac{\partial y}{\partial\lambda_1} & \frac{\partial y}{\partial\lambda_2} \\
         \frac{\partial z}{\partial\lambda_0} & \frac{\partial z}{\partial\lambda_1} & \frac{\partial z}{\partial\lambda_2}
    \end{smallmatrix}\right]\stackrel{\eqref{eq:barycentric_coors}}{=}
    \left[\begin{smallmatrix}
         x_0-x_3 & x_1-x_3 & x_2-x_3 \\
         y_0-y_3 & y_1-y_3 & y_2-y_3 \\
         z_0-z_3 & z_1-z_3 & z_2-z_3
    \end{smallmatrix}\right]
    .
    \label{eq:jacobian}
\end{equation}
Note that the relation of rectangular coordinates $(x,y,z)^\intercal$ and barycentric coordinates $(\lambda_0$, $\lambda_1$, $\lambda_2, \lambda_3)^\intercal$ of $\mathbf{p}$ is given in~\Cref{eq:barycentric_coors}, which shows that rectangular coordinates $(x,y,z)^\intercal$ of any point $\mathbf{p}$ are a linear combination of $\lambda_0$, $\lambda_1$, $\lambda_2$, and $\lambda_3$. Since all the four $\lambda_i$'s sum up to 1, we replace $\lambda_3$ by the other three. Thus, the derivatives in~\Cref{eq:jacobian} are all constants w.r.t. $\lambda_0$, $\lambda_1$, and $\lambda_2$, and the whole determinant of Jacobian matrix can be pulled out of the integral.  We further simplify the closed form of the remaining integral term and have
\begin{align}
    E(c)&=
    \frac{|J_c|}{60}\sum\nolimits_{i=0}^3\sum\nolimits_{j=i}^3(\hat{f}_i-f_i)(\hat{f}_j-f_j).
    \label{eq:integral_result}
\end{align}
Validation of CMSE is provided in Supplementary Materials.

\begin{table}[!th]
\scriptsize
    \centering
    \caption{Continuous versions of MSE, RMSE, NRMSE, and PSNR.}
    {\small
    \begin{tabular}{c|c}
        Metric & Definition \\ \hline
        CMSE & $\frac{1}{V_C}\sum_{i\in C}\iiint_{c_i} (\hat{f}(p(x,y,z))-f(p(x,y,z)))^2dxdydz$ \\
        CRMSE & $\sqrt{\texttt{CMSE}}$ \\
        CNRMSE & $\sqrt{\texttt{CMSE}}/(f_{\max}-f_{\min})$ \\
        CPSNR & $-20log_{10} \texttt{CNRMSE}$
    \end{tabular}
    }
    \label{tab:continuous_metrics}
\end{table}

%% file: sec-evaluation.tex
\begin{table}[!htb]
\scriptsize
    \centering
    \caption{Benchmark datasets (MPAS-O (hiRes) is in single-precision; all others are in double-precision.)}
    {\footnotesize
    \setlength{\tabcolsep}{1pt}
    \begin{tabular}{c|c|c|c|c}
        dataset & dim & attribute(s) & \# nodes & \# cells \\\hline
        synthetic & 2D & scalar & 1,000 & 1,903 \\
        MPAS-O & 2.5D & \makecell{temperature, salinity,\\ vorticity, velocities}
        & 235,055 & 478,834 \\
        \begin{tabular}{c}MPAS-O\\(hiRes)\end{tabular} & 2.5D & longWaveHeatFluxDown & 3,692,805 & 7,329,255 \\
        LES (small) & 3D & pressure, turb\_mu, velocities & 829,192 & 4,936,613 \\
        LES (large) & 3D & pressure, turb\_mu, velocities & 67,855,938 & 137,271,946 \\
        VFEM & 3D & pressure, velocity & 1,889,283 & 11,110,922 \\
        CAR & 3D & pressure, velocity & 8,565,665 & 49,419,394
    \end{tabular}
    }
    \label{tab:datasets}
\end{table}

\begin{figure*}[!th]
    \centering
    \includegraphics[width=\textwidth]{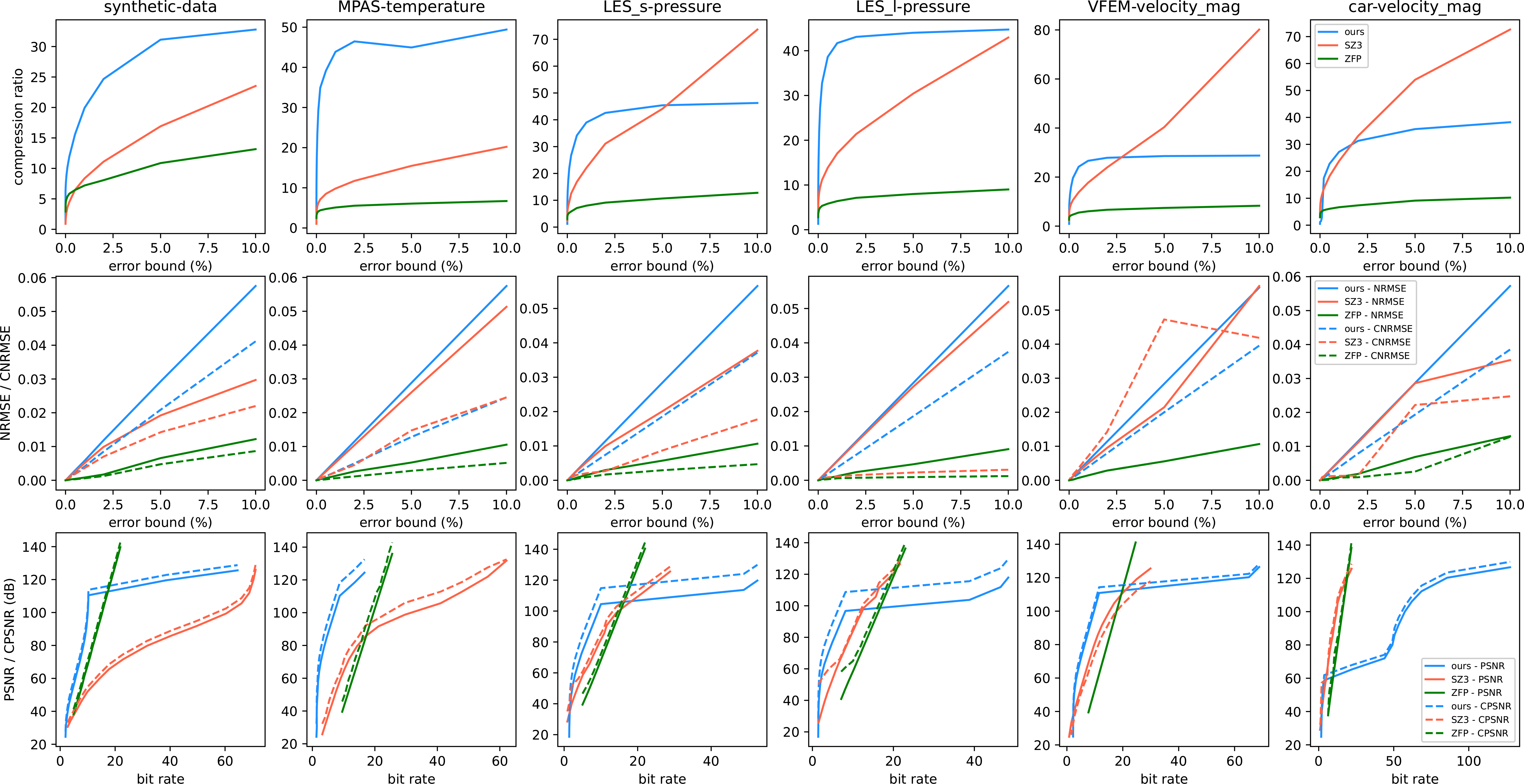}
    \caption{The performance of our method, SZ3, and ZFP on different datasets and their attributes.}
    \label{fig:performance}
\end{figure*}

\begin{figure}
    \centering
    \includegraphics[width=\linewidth]{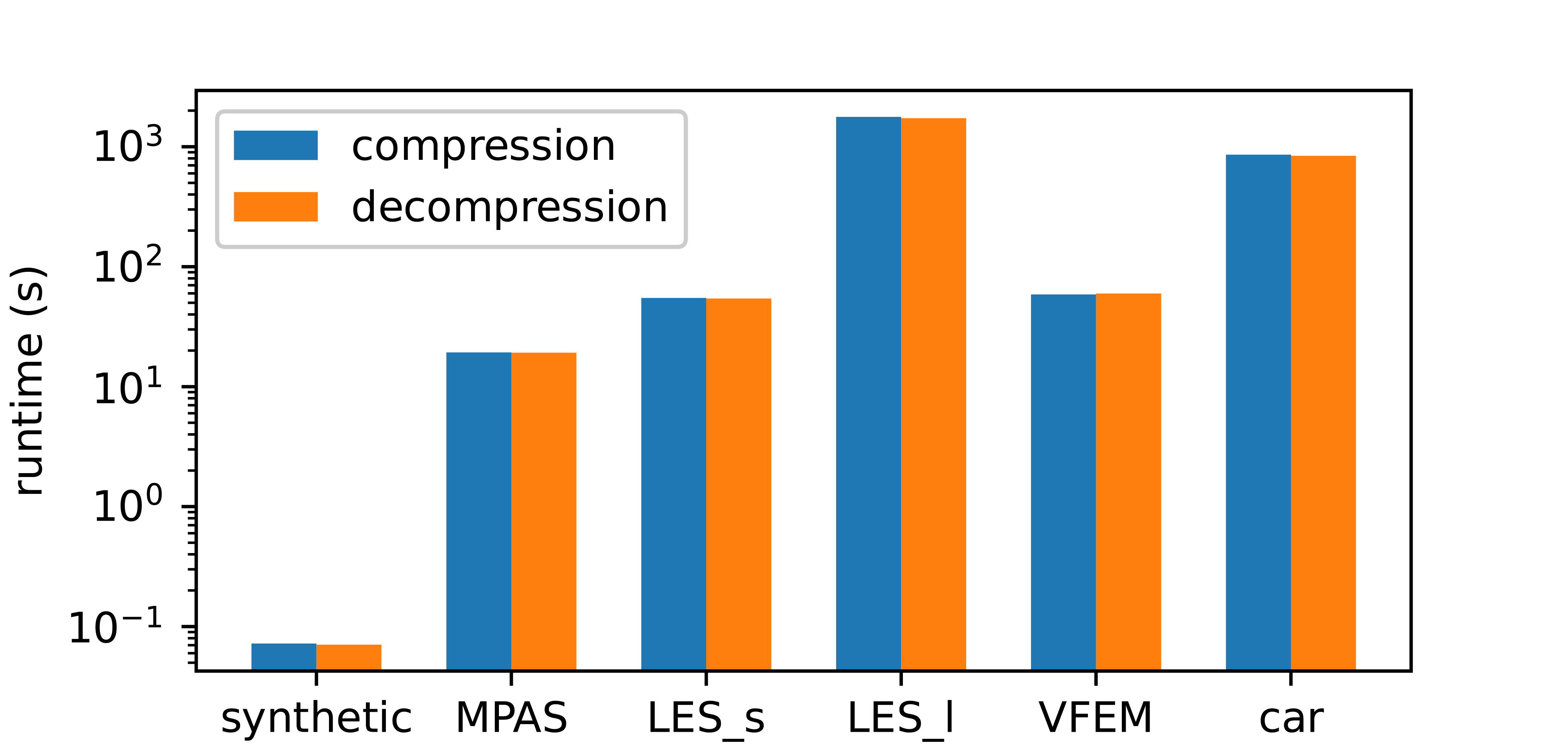}
    \caption{(a) Timings of (de)compression. They are almost consistent over attributes and $\xi$'s because the time complexity is always $O(|V|)$, so we average runtimes over all attributes and $\xi$'s for each dataset. 
    (b) Average time to process a vertex. Larger datasets need more time to process one node.}
    \label{fig:runtime}
\end{figure}

\begin{figure*}[!th]
    \centering
    \includegraphics[width=\textwidth]{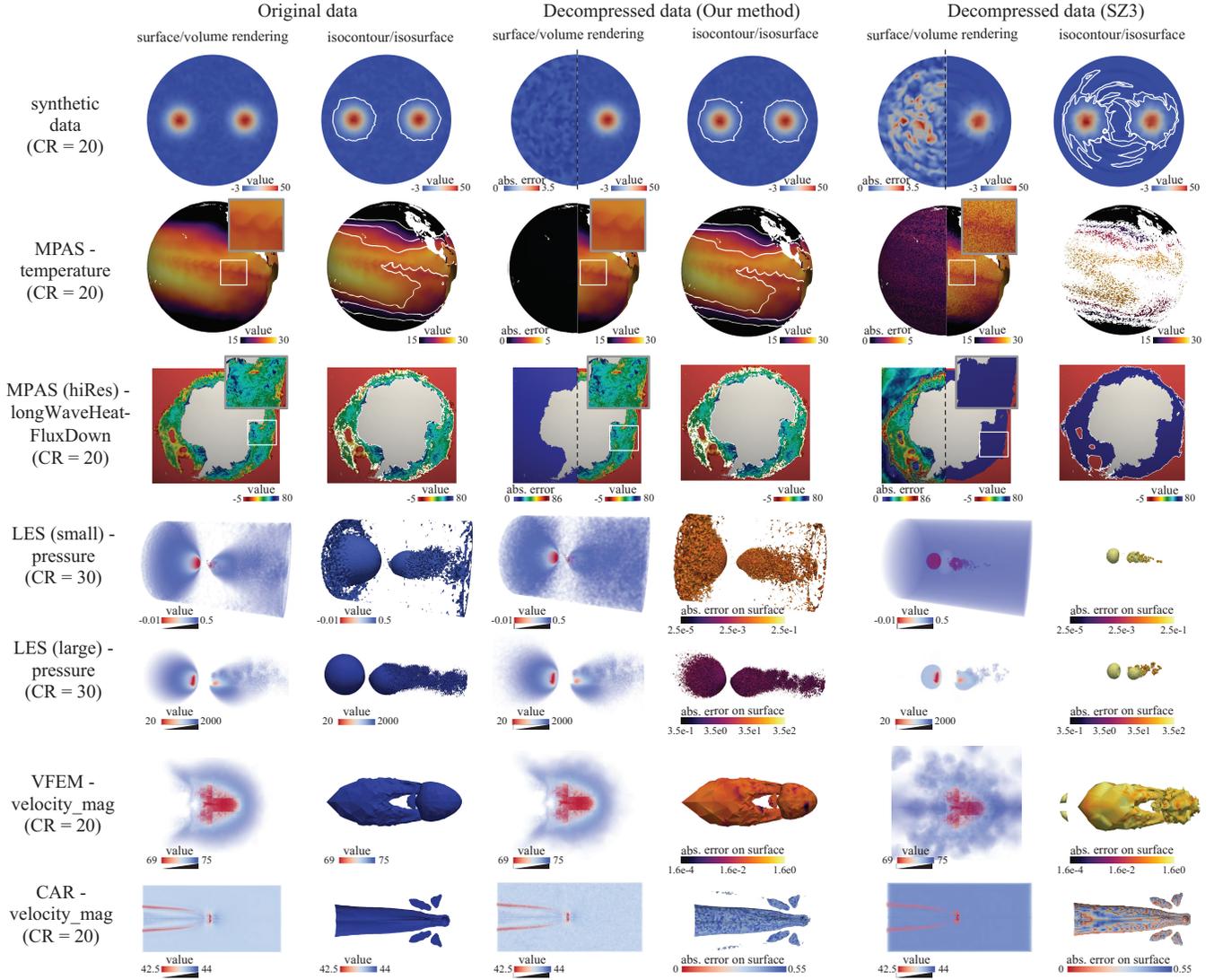}
    \caption{Volume rendering and isosurface results with the same compression ratio (CR) of our method and SZ3 on different datasets and their attributes. We have better rendering and isosurface results compared with SZ3 under the same CR for all datasets.}
    \label{fig:volRen}
\end{figure*}

\section{Evaluation}
\label{sec:evaluation}

This section compares our method with two state-of-the-art error-bounded lossy compressors, SZ3~\cite{liang2022sz3} and ZFP~\cite{lindstrom2014fixed}.
Note that SZ3 and ZFP are not originally designed for unstructured mesh data, and it is nontrivial to reorganize unstructured grid data into regular stencils (e.g., $4\times4\times4$ blocks used in ZFP) while keeping a high level of spatial coherence.  
The comparison aims to demonstrate how compression ratios and quality could significantly improve if spatial coherence were considered in our design.
Specifically, in our experiments, we treat nodal data as a 1D array directly from flattening the nodal values with their original in-memory ordering.
In addition to our baseline comparison in~\mbox{\Cref{sec:baseline}}, we further evaluate our method over a time-varying multivariate data (\mbox{\Cref{sec:timevarying}}) and demonstrate an ablation study in~\mbox{\Cref{sec:ablation}}, which compares the proposed method with SZ/ZFP when data is re-ordered by different traversal orders.
We prototype our compression and decompression algorithms using Python with no parallelization technology.  All experiments are based on a 2021 MacBook Pro with an Apple M1 CPU and 64 GB main memory. 

The specifications of our benchmark datasets are shown in~\Cref{tab:datasets}.  The synthetic data represents two Gaussian blobs with phase difference $\pi$ rotating around the center of a circular field.  The MPAS-O data are generated by the E3SM climate simulation~\cite{e3sm}. 
Mesh nodes in MPAS-O all have 3D coordinates but are located around the earth's surface and meshed by 2-polytopes, and thus are interpreted as 2.5D. 
MPAS-O has Voronoi mesh with hexagons and pentagons originally; we take its dual graph and derive a triangular mesh. Both LES-small and LES-large are from a large eddy simulation. VFEM and CAR are publicly available CFD datasets~\cite{cfd}.  

We evaluate our methods with four aspects: (1) size metrics vs. $\xi$~{(\%)}: compression ratio vs. $\xi$~{(\%)}, where $\xi$~{(\%)} is normalized $\xi$ in percentage scale by the range of original value. Bit rate is inversely proportional to compression ratio so we omit results of bit rate vs. $\xi$~{(\%)}; 
(2) error metrics vs. $\xi$~{(\%)}: NRMSE and CNRMSE vs. $\xi$~{(\%)}; 
(3) \textit{Rate-distortion} (i.e., error metrics vs. bit rate): PSNR and CPSNR vs. BR;
(4) Time metrics vs. $\xi$~{(\%)}: running times of compression and decompression vs. $\xi$~{(\%)}.

\subsection{Baseline Comparison}
\label{sec:baseline}

\textbf{Compression ratio vs. $\xi$~{(\%)}}. As shown in~\Cref{fig:performance}, compression ratio is the most significant advantage of our method compared with SZ3 and ZFP on most datasets tested by us, especially for lower error bounds, such as velocity magnitude on VFEM data. Also, the compression ratio by our method rises very fast as we relax $\xi$ when $\xi$ is small. For higher error bound, there might exist an intersection point of compression ratios by our method and SZ3 for some datasets (e.g., $\xi (\%)=3\%$ for VFEM and $\xi (\%)=1.8\%$ for CAR).

\begin{figure}[!th]
    \centering
    \includegraphics[width=\linewidth]{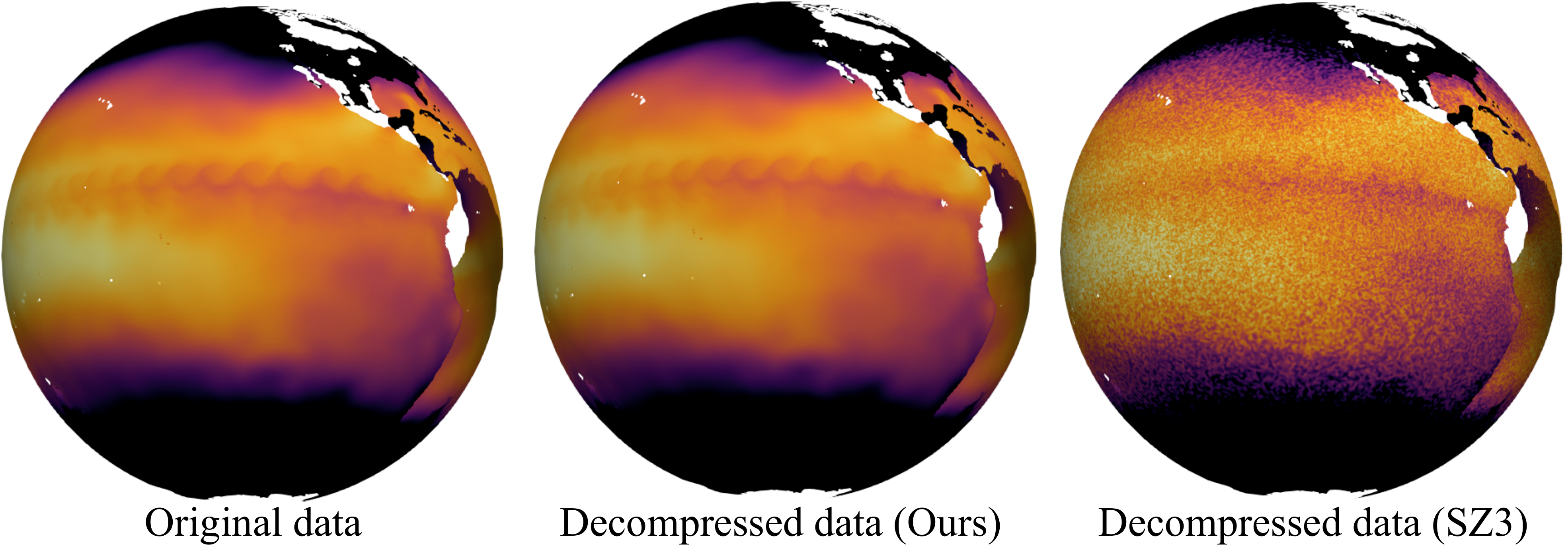}
    \caption{Surface rendering of original MPAS-O data and reconstructed data by our method and SZ3 under $\xi (\%)=5\%$. Rendering result of our method is very close to that of original data, but that of SZ3 has more noise-like dots and shows a different pattern around the equator compared to the other two rendering results. Our method has CNRMSE of 0.0128 and NRMSE of 0.0288; SZ3 has CNRMSE of 0.0148 and NRMSE of 0.0260. Our method provides better CMSE as well as visual performance but larger MSE, which manifests that CMSE measures visual quality better.}
    \label{fig:cmse_mpas}
\end{figure}

\textbf{NRMSE and CNRMSE vs. $\xi$~{(\%)}}. We show the results in Supplementary Materials. The performance ranking of our method and SZ3 is not consistent based on NRMSE and CNRMSE. For the temperature field of MPAS-O and velocity magnitude of CAR dataset around $\xi(\%)=5\%$, our method shows better CNRMSE than SZ3 does but higher NRMSE.
The difference in performance ranking on NRMSE and CNRMSE results from the mesh's geometry (as illustrated in~\Cref{sec:cmse}) and predictor. For example, MPAS-O data is sampled on the ocean only over the earth, while the land is not included. Although mesh nodes on MPAS-O data are almost uniform, the large number of mesh nodes on the boundary leads to different performances of the two methods on CMSE and MSE. We visualize the decompressed data by our method and SZ3 under $\xi (\%)=5\%$ in~\Cref{fig:cmse_mpas}, where our method has better CMSE with a better visual quality while SZ3 has better MSE.

\textbf{Rate distortion (PSNR and CPSNR vs. bit rate)}. Our method performs better according to rate distortion than SZ3 does on most datasets (except CAR) under low bit rate. The variations of PSNR and CPSNR are very close, so the ranking of coompressors hardly changes. ZFP shows the best performance under high bit rate.

\textbf{Running time}. The time complexity of our method is $O(|V|)$, independent to $\xi$. The plot of running time vs. $\xi (\%)$ verifies this statement and is provided in Supplementary Materials. We show the average running time over $\xi$ for the six datasets in~\Cref{fig:runtime} (a). 
We omit the comparison with SZ3 and ZFP as their C/C++ implementations deliver better performance than our Python implementation by nature.
The ratios of numbers of nodes to running times are in~\Cref{fig:runtime} (b). The relative difference of this ratio between datasets is significantly smaller than that of running time in~\Cref{fig:runtime} (a), but large datasets still take more time to process a node.

\textbf{Visual qualities}.  We compare both volume rendering and isosurface rendering of decompressed data by our method and SZ3 under the same compression ratio (\Cref{fig:volRen}).  
Note that visual comparisons with ZFP outputs at comparative compression ratios between 20 and 30 are unavailable for our benchmark datasets.
For most datasets, we have lower visual distortion compared to SZ3.  

\subsection{Evaluation with time-varying multivariate data}
\label{sec:timevarying}

\begin{figure}
    \centering
    \includegraphics[width=\linewidth]{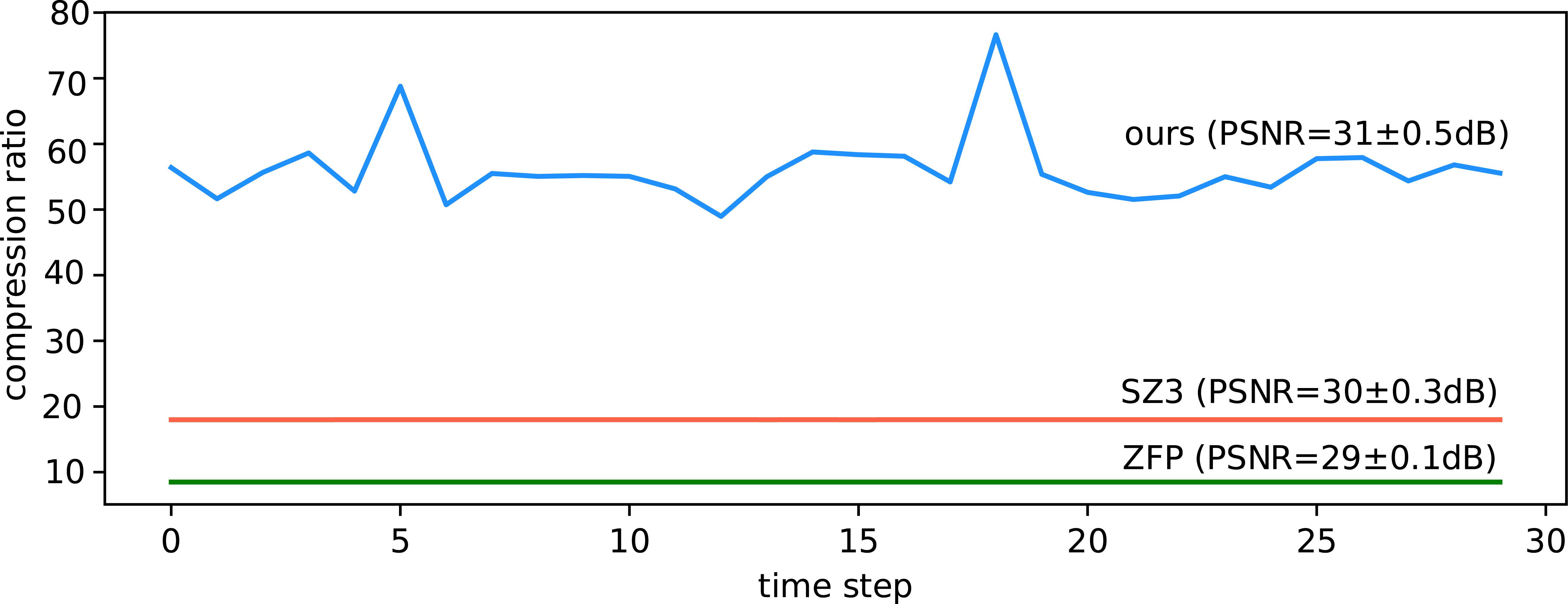}
    \caption{Compression ratios over time of (a) temperature and (b) meridional velocity on MPAS-O data under PSNR around 30 dB. The mesh structure takes 10.86 MB to store. We have a better compression ratio even under slightly better PSNR: Our method decreases the storage from 1.30 GB to 26.18 MB for temperature data and to 29.01 MB for meridional velocity data in 744 timesteps. SZ3 reduces 1.30 GB to 135.93 MB and 65.16 MB for temperature and meridional velocity, respectively, while ZFP reduces them to 261.75 MB and 167.23 MB. The compression ratio of our method slightly jitters because of the heuristics on seed placement.}
    \label{fig:MPAS_time_varying}
\end{figure}

As the mesh structure usually does not change over time or variable for time series multivariate data in many real-world applications such as ocean simulations, we evaluate our method with a time-series of MPAS-O data with two variables. In this case, the coordinates of nodes (three float64 points for each 3D point) and cell connectivity (three int32 points for three vertex indices of a cell) takes 10.86 MB to store. In contrast, nodal data (a float64 value for one variable of a point) in 744 timesteps of two variables takes 2.61 GB. 
We evaluate the compression ratio over time of temperature on the MPAS-O dataset under very close PSNRs (\Cref{fig:MPAS_time_varying}). After compression, the storage that one needs to save nodal data of 744 timesteps is reduced to 55.19 MB, compared with 201.09 MB by SZ3 and 428.98 MB by ZFP.

\subsection{Ablation Studies}
\label{sec:ablation}

We conduct additional experiments to support the hypothesis that (H1) our coupled barycentric prediction and traversal scheme effectively compress nodal data.  While it is impractical to disentangle prediction and traversal entirely, we design experiments to explore whether reordering alone is sufficient to compress nodal data effectively.  We also explore (H2) whether alternative traversal ordering schemes could improve compression.  For H1, we impose off-the-shelf lossy compressors over our reordered data to see if SZ3/ZFP over reordered data (noted as SZ3/ZFP-over-ours) outperforms our coupled method.  To test H2, we explored four alternative strategies:
\begin{itemize}
    \item Descending strategy (implicit ordering used as the default in our method): each sequence traversing towards the neighbor cell with the minimum storage index;
    \item Ascending strategy (implicit ordering): each sequence traversing towards the neighbor cell with the maximum storage index;
    \item Mixed strategy (implicit ordering): each sequence randomly uses either the descending or ascending strategy;
    \item Accuracy-prioritized strategy (explicit ordering): each sequence traversing towards the next neighboring cell with minimum prediction error on the new vertex, incurring minimum prediction error but requiring explicitly storing the vertex ordering.
\end{itemize}

\begin{figure}
    \centering
    \includegraphics[width=\linewidth]{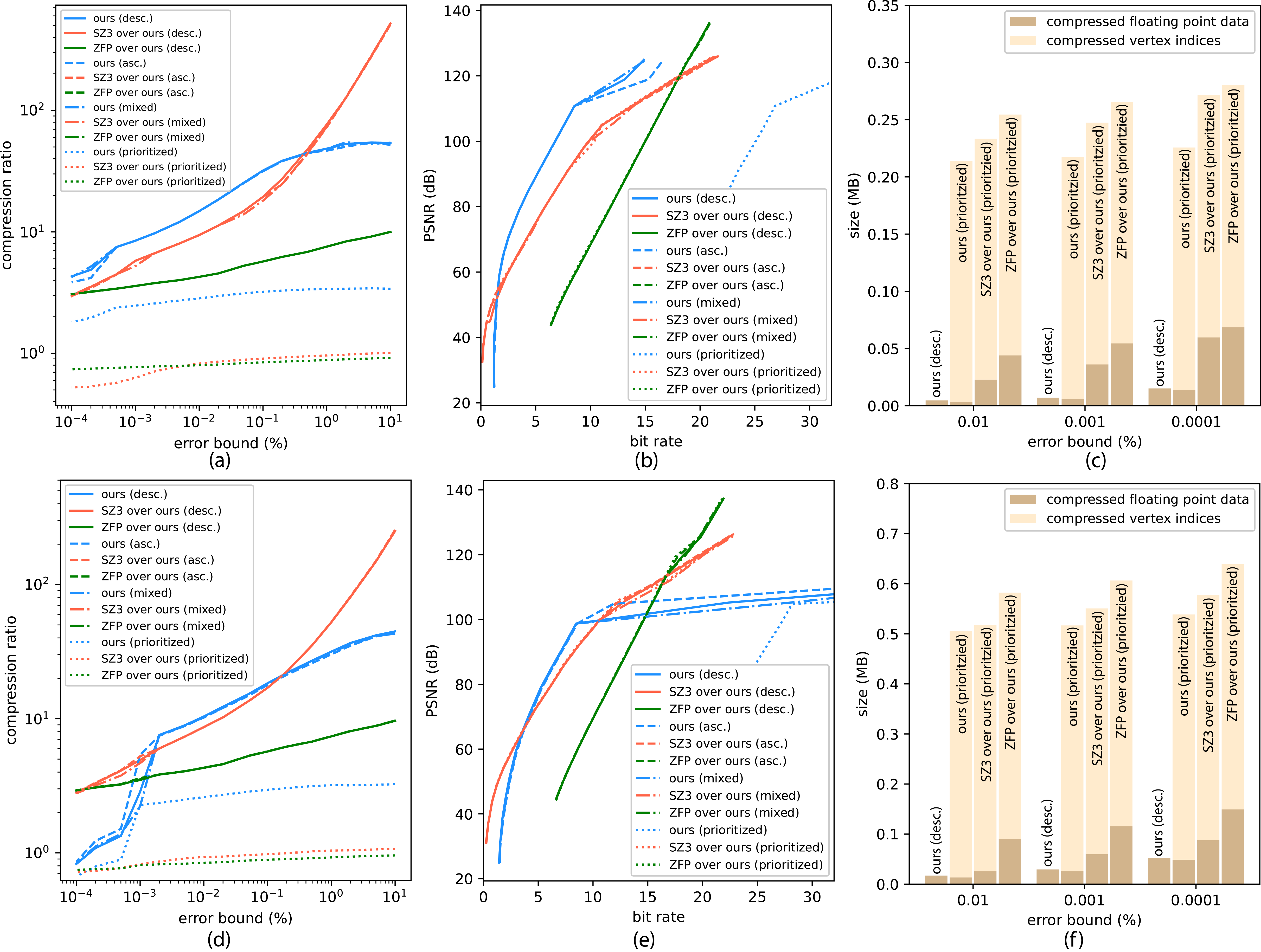}
    \caption{Ablation study with temperature in the MPAS-O dataset (first row) and velocity magnitude in the LES (small) dataset (second row). Especially, (c) and (f) illustrate the decomposition of floating-point data and vertex ordering in the compressed file size for the accuracy-prioritized strategy.}
    \label{fig:ablation}
\end{figure}

Observation 1: While our reordering scheme alone (SZ3/ZFP-over-ours) already improves off-the-shelf compressors, our coupled prediction-traversal compressor further outperforms SZ3/ZFP-over-ours to different extents.  
Note that this is not a strictly controlled experiment because SZ3/ZFP features diverse prediction/transformation schemes for the reordered data.  Still, our combined prediction-traversal scheme outperforms SZ3/ZFP-over-ours to different extents depending on error bounds and data (comparisons in~\mbox{\Cref{fig:ablation}} a/d and b/e), supporting H1 that both our prediction and reordering contribute to the effective compression.

Observation 2: Among the alternative traversal orderings, the three implicit ordering strategies are similar, but the explicit ordering scheme (although prioritizing accuracy during traversal) is far inferior to implicit orderings.  As shown in the left two columns of~\mbox{\Cref{fig:ablation}}, the three implicit strategies exhibit the same trend on compression ratio and rate-distortion (H2).  Besides, in the accuracy-prioritized traversal, the overhead of additionally stored vertex ordering dominates compressed outputs (up to 99\%).

%% file: sec-discussion.tex
\section{Limitations}
\label{sec:discussion}

Although this work is not limited to 2D and 3D simplicial meshes as we used in~\Cref{sec:evaluation}, there are several unexplored topics. First, we cover neither simplicial meshes beyond 3D nor other unstructured data in 2D and 3D. Also, different triangulations lead to different results using different interpolation methods.
Whether arbitrary triangulation ways would show similar results or if interpolation in arbitrary polytope is required is still an open question.

One limitation is the flattened compression ratio as $\xi$ increases.  This is because the predicted value on every new vertex accumulates errors from all the vertices of its adjacent cell; as $\xi$ increases, the prediction error could climb fast, leading to higher entropy/complexity of quantization codes and damaging the compression ratio.  
Meanwhile, the storage of seed/sequence will also incur overhead.
Likewise, our method also exhibits a turning point in rate-distortion curves, caused by the sudden increase of sequences as $\xi$ decreases.
As an example, the velocity magnitude of the VFEM dataset needs 58,130 sequences for $\xi$(\%)=0.0002\% but only 1,853 sequences for $\xi$(\%)=0.0005\%.
As such, there is a minor improvement in PSNR (because of the small error bound change) with a suddenly increased bit rate, leading to the flattening effect.

Another limitation in the current algorithm is parallelization.  The time complexity of our algorithm is $O(|V|)$ because every node is visited once. However, one needs to wait until the traversal starting with the current seed terminates to begin the next traversal. 
It would be possible to select multiple seeds at first to start traversal, but high-frequency communication among traversals is required to exchange information on visited nodes, and running time may not benefit from the communication.

%% file: sec-conclusions.tex
\section{Conclusions and future work}
\label{sec:conclusion}

We introduce an error-bounded lossy compressor for unstructured mesh data.
The key to our compression is to use a barycentric extrapolation scheme to predict nodal values paired with a greedy mesh traversal strategy to represent nodal values as separate sequences.  Each sequence starts with a seed cell and iteratively grows with an implicit traversal order until the newly added cell cannot be predicted with the error-controlled quantizer.
We also introduce a metric that evaluates data distortion in continuous domains, CMSE, a more general version of MSE tailored for unstructured meshes.
We evaluate our algorithm on several datasets and demonstrate superior quality and compression ratio compared with state-of-the-art error-bounded lossy compressors agnostic to unstructured meshes.

This work can be extended in multiple ways. First, one can use our strategy to predict point cloud data, which is easy to triangulate by methods such as Delaunay triangulation. Second, there might be metrics to precheck if our method or other compressors, such as SZ3 and ZFP, can achieve a better compression ratio without executing all the compression methods. Third, further research is needed to understand the impact of lossy compression on unstructured grid volume visualization.

%% file: main.bbl
\newcommand{\etalchar}[1]{$^{#1}$}
\begin{thebibliography}{\uppercase{ATWK20}}

\bibitem[ATWK20]{ainsworth2020multilevel}
\textsc{Ainsworth M., Tugluk O., Whitney B., Klasky S.}:
\newblock Multilevel techniques for compression and reduction of scientific data---the unstructured case.
\newblock \emph{SIAM Journal on Scientific Computing 42}, 2 (2020), A1402--A1427.

\bibitem[BRLP19]{ballester2019tthresh}
\textsc{Ballester-Ripoll R., Lindstrom P., Pajarola R.}:
\newblock {TTHRESH}: Tensor compression for multidimensional visual data.
\newblock \emph{IEEE Transactions on Visualization and Computer Graphics 26}, 9 (2019), 2891--2903.

\bibitem[BW08]{bachthaler2008continuous}
\textsc{Bachthaler S., Weiskopf D.}:
\newblock Continuous scatterplots.
\newblock \emph{IEEE Transactions on Visualization and Computer Graphics 14}, 6 (2008), 1428--1435.

\bibitem[CCM{\etalchar{*}}00]{cignoni2000simplification}
\textsc{Cignoni P., Costanza D., Montani C., Rocchini C., Scopigno R.}:
\newblock Simplification of tetrahedral meshes with accurate error evaluation.
\newblock In \emph{Proceedings IEEE Visualization} (2000), pp.~85--92.

\bibitem[Cho97]{chow1997optimized}
\textsc{Chow M.~M.}:
\newblock \emph{Optimized geometry compression for real-time rendering}.
\newblock 1997.

\bibitem[CL03]{chiang2003progressive}
\textsc{Chiang Y.-J., Lu X.}:
\newblock Progressive simplification of tetrahedral meshes preserving all isosurface topologies.
\newblock \emph{Computer Graphics Forum 22}, 3 (2003), 493--504.

\bibitem[CMS98]{cignoni1998comparison}
\textsc{Cignoni P., Montani C., Scopigno R.}:
\newblock A comparison of mesh simplification algorithms.
\newblock \emph{Computers \& Graphics 22}, 1 (1998), 37--54.

\bibitem[DCM12]{duffy2012integrating}
\textsc{Duffy B., Carr H., M{\"o}ller T.}:
\newblock Integrating isosurface statistics and histograms.
\newblock \emph{IEEE Transactions on Visualization and Computer Graphics 19}, 2 (2012), 263--277.

\bibitem[DLRS10]{de2010triangulations}
\textsc{De~Loera J., Rambau J., Santos F.}:
\newblock \emph{Triangulations: Structures for Algorithms and Applications}.
\newblock Springer, 2010.

\bibitem[EMX02]{estkowski2002optimal}
\textsc{Estkowski R., Mitchell J.~S., Xiang X.}:
\newblock Optimal decomposition of polygonal models into triangle strips.
\newblock In \emph{Proceedings of the 18th Annual Symposium on Computational Geometry} (2002), pp.~254--263.

\bibitem[GSG96]{gross1996efficient}
\textsc{Gross M.~H., Staadt O.~G., Gatti R.}:
\newblock Efficient triangular surface approximations using wavelets and quadtree data structures.
\newblock \emph{IEEE Transactions on Visualization and Computer Graphics 2}, 2 (1996), 130--143.

\bibitem[Huf52]{huffman1952method}
\textsc{Huffman D.~A.}:
\newblock A method for the construction of minimum-redundancy codes.
\newblock \emph{Proceedings of the IRE 40}, 9 (1952), 1098--1101.

\bibitem[HZCW22]{Han-VIS21}
\textsc{Han J., Zheng H., Chen D.~Z., Wang C.}:
\newblock {STNet}: An end-to-end generative framework for synthesizing spatiotemporal super-resolution volumes.
\newblock \emph{IEEE Transactions on Visualization and Computer Graphics 28}, 1 (2022), 270--280.

\bibitem[IKK12]{Iverson12}
\textsc{Iverson J., Kamath C., Karypis G.}:
\newblock Fast and effective lossy compression algorithms for scientific datasets.
\newblock In \emph{Euro-Par 2012 Parallel Processing} (Berlin, Heidelberg, 2012), Kaklamanis C., Papatheodorou T., Spirakis P.~G., (Eds.), Springer Berlin Heidelberg, pp.~843--856.

\bibitem[ILRS03]{ibarria2003out}
\textsc{Ibarria L., Lindstrom P., Rossignac J., Szymczak A.}:
\newblock Out-of-core compression and decompression of large n-dimensional scalar fields.
\newblock \emph{Computer Graphics Forum 22(3)} (2003), 343--348.

\bibitem[Kam20]{kamath2020compressing}
\textsc{Kamath C.}:
\newblock Compressing unstructured mesh data from simulations using machine learning.
\newblock \emph{International Journal of Data Science and Analytics 9}, 1 (2020), 113--130.

\bibitem[LBTM20]{e3sm}
\textsc{Leung L.~R., Bader D.~C., Taylor M.~A., McCoy R.~B.}:
\newblock An introduction to the {E3SM} special collection: Goals, science drivers, development, and analysis.
\newblock \emph{Journal of Advances in Modeling Earth Systems 12}, 11 (2020), e2019MS001821.

\bibitem[LDC{\etalchar{*}}23]{LiangDCRLOCPG23}
\textsc{Liang X., Di S., Cappello F., Raj M., Liu C., Ono K., Chen Z., Peterka T., Guo H.}:
\newblock Toward feature-preserving vector field compression.
\newblock \emph{IEEE Transactions on Visualization and Computer Graphics 29}, 12 (2023), 5434--5450.

\bibitem[LDT{\etalchar{*}}18]{liang2018error}
\textsc{Liang X., Di S., Tao D., Li S., Li S., Guo H., Chen Z., Cappello F.}:
\newblock Error-controlled lossy compression optimized for high compression ratios of scientific datasets.
\newblock In \emph{Proceedings of 2018 IEEE International Conference on Big Data (Big Data)} (2018), pp.~438--447.

\bibitem[LDZ{\etalchar{*}}21]{liu2021exploring}
\textsc{Liu J., Di S., Zhao K., Jin S., Tao D., Liang X., Chen Z., Cappello F.}:
\newblock Exploring autoencoder-based error-bounded compression for scientific data.
\newblock In \emph{Proceedings of 2021 IEEE International Conference on Cluster Computing (CLUSTER)} (2021), pp.~294--306.

\bibitem[LI06]{lindstrom2006fast}
\textsc{Lindstrom P., Isenburg M.}:
\newblock Fast and efficient compression of floating-point data.
\newblock \emph{IEEE Transactions on Visualization and Computer Graphics 12}, 5 (2006), 1245--1250.

\bibitem[Lin14]{lindstrom2014fixed}
\textsc{Lindstrom P.}:
\newblock Fixed-rate compressed floating-point arrays.
\newblock \emph{IEEE Transactions on Visualization and Computer Graphics 20}, 12 (2014), 2674--2683.

\bibitem[LJLB21]{Lu21}
\textsc{Lu Y., Jiang K., Levine J.~A., Berger M.}:
\newblock Compressive neural representations of volumetric scalar fields.
\newblock \emph{Computer Graphics Forum 40}, 3 (2021), 135--146.

\bibitem[LSE{\etalchar{*}}11]{lakshminarasimhan2011compressing}
\textsc{Lakshminarasimhan S., Shah N., Ethier S., Klasky S., Latham R., Ross R., Samatova N.~F.}:
\newblock Compressing the incompressible with isabela: In-situ reduction of spatio-temporal data.
\newblock In \emph{Euro-Par 2011 Parallel Processing: 17th International Conference, Euro-Par 2011, Bordeaux, France, August 29-September 2, 2011, Proceedings, Part I 17} (2011), Springer, pp.~366--379.

\bibitem[LSO{\etalchar{*}}17a]{li2017spatiotemporal}
\textsc{Li S., Sane S., Orf L., Mininni P., Clyne J., Childs H.}:
\newblock Spatiotemporal wavelet compression for visualization of scientific simulation data.
\newblock In \emph{Proceedings of 2017 IEEE International Conference on Cluster Computing (CLUSTER)} (2017), pp.~216--227.

\bibitem[LSO{\etalchar{*}}17b]{LiSOMCC17}
\textsc{Li S., Sane S., Orf L., Mininni P.~D., Clyne J.~P., Childs H.}:
\newblock Spatiotemporal wavelet compression for visualization of scientific simulation data.
\newblock In \emph{Proceedings of 2017 {IEEE} International Conference on Cluster Computing, {CLUSTER}} (2017), pp.~216--227.

\bibitem[LZD{\etalchar{*}}22]{liang2022sz3}
\textsc{Liang X., Zhao K., Di S., Li S., Underwood R., Gok A.~M., Tian J., Deng J., Calhoun J.~C., Tao D., et~al.}:
\newblock {SZ}3: A modular framework for composing prediction-based error-bounded lossy compressors.
\newblock \emph{IEEE Transactions on Big Data 9}, 2 (2022), 485--498.

\bibitem[MLL{\etalchar{*}}21]{Martel21}
\textsc{Martel J. N.~P., Lindell D.~B., Lin C.~Z., Chan E.~R., Monteiro M., Wetzstein G.}:
\newblock {ACORN}: Adaptive coordinate networks for neural scene representation.
\newblock \emph{ACM Transactions on Graphics 40}, 4 (2021), 58:1--58:13.

\bibitem[NE04]{natarajan2004simplification}
\textsc{Natarajan V., Edelsbrunner H.}:
\newblock Simplification of three-dimensional density maps.
\newblock \emph{IEEE Transactions on Visualization and Computer Graphics 10}, 5 (2004), 587--597.

\bibitem[PKJ05]{Peng05}
\textsc{Peng J., Kim C.-S., {Jay Kuo} C.-C.}:
\newblock Technologies for {3D} mesh compression: A survey.
\newblock \emph{Journal of Visual Communication and Image Representation 16}, 6 (2005), 688--733.

\bibitem[RB93]{rossignac1993multi}
\textsc{Rossignac J., Borrel P.}:
\newblock Multi-resolution {3D} approximations for rendering complex scenes.
\newblock In \emph{Modeling in Computer Graphics: Methods and Applications}. Springer, 1993, pp.~455--465.

\bibitem[RL00]{rusinkiewicz2000qsplat}
\textsc{Rusinkiewicz S., Levoy M.}:
\newblock {QSplat}: A multiresolution point rendering system for large meshes.
\newblock In \emph{Proceedings of the 27th Annual Conference on Computer graphics and Interactive Techniques} (2000), pp.~343--352.

\bibitem[SFH{\etalchar{*}}18]{salloum2018optimal}
\textsc{Salloum M., Fabian N.~D., Hensinger D.~M., Lee J., Allendorf E.~M., Bhagatwala A., Blaylock M.~L., Chen J.~H., Templeton J.~A., Tezaur I.}:
\newblock Optimal compressed sensing and reconstruction of unstructured mesh datasets.
\newblock \emph{Data Science and Engineering 3} (2018), 1--23.

\bibitem[Si19]{si2019simple}
\textsc{Si H.}:
\newblock A simple algorithm to triangulate a special class of {3D} non-convex polyhedra without {Steiner} points.
\newblock In \emph{Numerical Geometry, Grid Generation and Scientific Computing} (2019), Springer, pp.~61--71.

\bibitem[SKJ{\etalchar{*}}20]{salloum2020comparing}
\textsc{Salloum M., Karlson K.~N., Jin H., Brown J.~A., Bolintineanu D.~S., Long K.~N.}:
\newblock Comparing field data using {Alpert} multi-wavelets.
\newblock \emph{Computational Mechanics 66} (2020), 893--910.

\bibitem[SMB{\etalchar{*}}20]{Sitzmann-NIPS20}
\textsc{Sitzmann V., Martel J. N.~P., Bergman A.~W., Lindell D.~B., Wetzstein G.}:
\newblock Implicit neural representations with periodic activation functions.
\newblock In \emph{Proceedings of Advances in Neural Information Processing Systems} (2020).

\bibitem[SPCT18]{soler2018topologically}
\textsc{Soler M., Plainchault M., Conche B., Tierny J.}:
\newblock Topologically controlled lossy compression.
\newblock In \emph{Proceedings of 2018 IEEE Pacific Visualization Symposium (PacificVis)} (2018), pp.~46--55.

\bibitem[SW03]{schneider2003compression}
\textsc{Schneider J., Westermann R.}:
\newblock Compression domain volume rendering.
\newblock In \emph{Proceedings of IEEE Visualization 2003} (2003), pp.~293--300.

\bibitem[TDCC17]{tao2017significantly}
\textsc{Tao D., Di S., Chen Z., Cappello F.}:
\newblock Significantly improving lossy compression for scientific data sets based on multidimensional prediction and error-controlled quantization.
\newblock In \emph{Proceedings of 2017 IEEE International Parallel and Distributed Processing Symposium (IPDPS)} (2017), IEEE, pp.~1129--1139.

\bibitem[Tri]{cfd}
\textsc{Tricoche X.}:
\newblock https://www.cs.purdue.edu/homes/cs530/projects/data/final/cfd.

\bibitem[UBF{\etalchar{*}}05]{uesu2005simplification}
\textsc{Uesu D., Bavoil L., Fleishman S., Shepherd J., Silva C.~T.}:
\newblock Simplification of unstructured tetrahedral meshes by point sampling.
\newblock In \emph{Proceedings of the Fourth International Workshop on Volume Graphics} (2005), IEEE, pp.~157--238.

\bibitem[WGS{\etalchar{*}}23]{WursterGSPX23}
\textsc{Wurster S.~W., Guo H., Shen H., Peterka T., Xu J.}:
\newblock Deep hierarchical super resolution for scientific data.
\newblock \emph{IEEE Transactions of Visualization and Computer Graphics 29}, 12 (2023), 5483--5495.

\bibitem[WHW22]{WeissHW22}
\textsc{Weiss S., Herm{\"{u}}ller P., Westermann R.}:
\newblock Fast neural representations for direct volume rendering.
\newblock \emph{Comput. Graph. Forum 41}, 6 (2022), 196--211.

\bibitem[WPG{\etalchar{*}}23]{WangPGTJTSD0FLC23}
\textsc{Wang D., Pulido J., Grosset P., Tian J., Jin S., Tang H., Sexton J.~M., Di S., Zhao K., Fang B., Lukic Z., Cappello F., Ahrens J.~P., Tao D.}:
\newblock {AMRIC:} {A} novel in situ lossy compression framework for efficient {I/O} in adaptive mesh refinement applications.
\newblock In \emph{Proceedings of the International Conference for High Performance Computing, Networking, Storage and Analysis, {SC} 2023, Denver, CO, USA, November 12-17, 2023} (2023), Arnold D., Badia R.~M., Mohror K.~M., (Eds.), {ACM}, pp.~44:1--44:15.

\bibitem[XTS{\etalchar{*}}22]{INR22}
\textsc{Xie Y., Takikawa T., Saito S., Litany O., Yan S., Khan N., Tombari F., Tompkin J., Sitzmann V., Sridhar S.}:
\newblock Neural fields in visual computing and beyond.
\newblock \emph{Computer Graphics Forum} (2022).

\bibitem[YLGW24]{yan2023toposz}
\textsc{Yan L., Liang X., Guo H., Wang B.}:
\newblock {TopoSZ}: Preserving topology in error-bounded lossy compression.
\newblock \emph{IEEE Transactions on Computer Graphics and Visualization 30}, 1 (2024), 1302--1312.

\bibitem[ZDD{\etalchar{*}}21]{zhao2021optimizing}
\textsc{Zhao K., Di S., Dmitriev M., Tonellot T.-L.~D., Chen Z., Cappello F.}:
\newblock Optimizing error-bounded lossy compression for scientific data by dynamic spline interpolation.
\newblock In \emph{Proceedings of 2021 IEEE 37th International Conference on Data Engineering (ICDE)} (2021), IEEE, pp.~1643--1654.

\bibitem[ZDL{\etalchar{*}}20]{zhao2020significantly}
\textsc{Zhao K., Di S., Liang X., Li S., Tao D., Chen Z., Cappello F.}:
\newblock Significantly improving lossy compression for {HPC} datasets with second-order prediction and parameter optimization.
\newblock In \emph{Proceedings of the 29th International Symposium on High-Performance Parallel and Distributed Computing} (2020), pp.~89--100.

\bibitem[ZGS{\etalchar{*}}22]{ZhangGSWP22}
\textsc{Zhang Y., Guo H., Shang L., Wang D., Peterka T.}:
\newblock A multi-branch decoder network approach to adaptive temporal data selection and reconstruction for big scientific simulation data.
\newblock \emph{{IEEE} Trans. Big Data 8}, 6 (2022), 1637--1649.

\bibitem[{ZST}]{zstd}
\textsc{{ZSTD}}:
\newblock http://www.zstd.net.

\end{thebibliography}
